\documentclass[twocolumn,showpacs,preprintnumbers,prl]{revtex4}
\usepackage{graphicx,graphics,epsfig,amsmath,color}

\newcommand{\sbL}{\bar{s}_{\scriptscriptstyle L}}
\newcommand{\cbL}{\bar{c}_{\scriptscriptstyle L}}
\newcommand{\bR}{b_{\scriptscriptstyle R}}
\newcommand{\bL}{b_{\scriptscriptstyle L}}
\newcommand{\cL}{c_{\scriptscriptstyle L}}
\newcommand{\ord}{{\cal{O}}}
\newcommand{\as}{\alpha_s}
\newcommand{\BtoXsgamma}{\bar{B} \to X_s \gamma} 

\newcommand{\mb}{m_b}
\newcommand{\mt}{m_t}
\newcommand{\MW}{M_{\scriptscriptstyle W}}
\newcommand{\MZ}{M_{\scriptscriptstyle Z}}
\newcommand{\GeV}{{\rm \ GeV}}
\newcommand{\btosgamma}{b \to s \gamma} 
\newcommand{\Leff}{{\cal{L}}_{\rm eff}}
\newcommand{\LQCDQED}{{\cal{L}}_{{\rm QCD} \times {\rm QED}}} 
\newcommand{\SUC}{SU (3)_C}
\newcommand{\beq}{\begin{equation}}
\newcommand{\eeq}{\end{equation}}
\newcommand{\f}{\frac}
\newcommand{\gs}{g_s}

\newcommand{\btosgluon}{b \to s g}
\newcommand{\Fig}[1]{FIG. \ref{#1}}

\newcommand{\Dsl}{D \hspace{-.70em} / \hspace{.25em}}
\newcommand{\smallDsl}{D \hspace{-.50em} / \hspace{.1em}}
\newcommand{\Gsl}{G \hspace{-.70em} / \hspace{.25em}}
\newcommand{\smallGsl}{G \hspace{-.50em} / \hspace{.1em}}
\newcommand{\NF}{n_f}
\newcommand{\ND}{n_d}
\newcommand{\NU}{n_u}
\newcommand{\MSbar}{\overline{\rm MS}}
\newcommand{\zetathree}{\zeta (3)}
\newcommand{\btos}{b \to s} 
\newcommand{\mub}{\mu_b}
\newcommand{\muw}{\mu_{\scriptscriptstyle W}}
\newcommand{\Eq}[1]{Eq.~(\ref{#1})}
\newcommand{\non}{\nonumber}

\begin{document}

\allowdisplaybreaks

\preprint{IPPP/05/15; DCPT/05/30; FERMILAB-Pub-05-076-T; IFT-9/2005;
arXiv:hep-ph/0504194} 

\title{
\boldmath
Three-Loop Mixing of Dipole Operators 
\unboldmath}
\author{Martin~Gorbahn${}^1$, Ulrich~Haisch${}^2$ and Miko{\l}aj
Misiak${}^3$}  

\affiliation{
$^1\!\!\!$ IPPP, Physics Department, University of Durham, DH1 3LE,
Durham, UK \\  
$^2\!\!\!$ Theoretical Physics Department, Fermilab, Batavia, IL60510,
USA \\ 
$^3\!\!\!$ Institute of Theoretical Physics, Warsaw University, 00-681
Warsaw, Poland 
}

\date{August 12, 2005}

\begin{abstract}
\noindent

We calculate the complete three-loop $\ord (\as^3)$ anomalous
dimension matrix for the dimension-five dipole operators that arise in
the Standard Model after integrating out the top quark and the heavy
electroweak bosons. Our computation completes the three-loop anomalous
dimension matrix of operators that govern low-energy $| \Delta F | =
1$ flavor-changing processes, and represents an important ingredient
of the next-to-next-to-leading order QCD analysis of the $\BtoXsgamma$
decay.    

\end{abstract}

\maketitle

Weak interaction phenomena at energies much below the electroweak
scale are most conveniently described in the framework of an effective
theory that is derived from the Standard Model (SM) by integrating out
the top quark and the heavy electroweak  bosons. The
Lagrangian of such an effective theory 
\beq \label{eq:leff.generic}
\Leff = \LQCDQED + {\cal L}_{\rm weak} \, , 
\eeq
is a sum of the conventional QCD$\times$QED Lagrangian for the
remaining SM fields and a linear combination  
\beq \label{eq:l.weak}
{\cal L}_{\rm weak} \propto \sum_i C_i(\mu) Q_i \, , 
\eeq
of dimension $\geq5$ operators $Q_i$ that are built out of those light
fields. $C_i (\mu)$ are the corresponding Wilson coefficients that
depend on the renormalization scale $\mu$.

For most phenomenological applications, only operators of dimensions 
five and six are relevant. The complete set of such operators consists
of: {\it i)} dimension-six four-fermion operators, {\it ii)}
dimension-six purely gluonic operators, and {\it iii)} dimension-five
dipole operators  
\beq \label{eq:dipoles}
\bar{\psi} \sigma^{\mu\nu} \psi' F_{\mu\nu} \, , 
\hspace{1cm} {\rm and} \hspace{1cm} 
\bar{\psi} \sigma^{\mu\nu} T^a \psi' G^a_{\mu\nu} \, . 
\eeq
Here $\sigma^{\mu\nu} = i/2 \, [ \gamma^\mu, \gamma^\nu ]$, while
$\psi$ and $\psi'$ stand for fermion fields of opposite
chiralities. Their flavors may or may not be the same. The
electromagnetic and strong field strength tensors are denoted by
$F_{\mu \nu}$ and $G_{\mu\nu}^a$, respectively. $T^a$ are the $\SUC$
generators for the considered fermions. For off-shell calculations,
additional operators that vanish by the QCD$\times$QED Equations of
Motion (EOM) must be included (see below). 

The structure of ${\cal L}_{\rm weak}$ remains the same in any $\SUC 
\times U(1)_{\rm em}$ gauge-invariant extension of the SM that
does not contain exotic light bosons. New physics effects are
therefore encoded in the values of the Wilson coefficients only.  

The dipole operators introduced in \Eq{eq:dipoles} are relevant
in a  variety of phenomenological applications, ranging from electric
and magnetic moments of the leptons and nucleons to radiative decays,
such as $\mu \to e \gamma$, $\Omega \to \Xi \gamma$, $B \to K^*
\gamma$, $B \to \rho \gamma$ and $\BtoXsgamma$.    

The Wilson coefficients are determined by matching Green's functions
of the effective theory and the SM (or its extension) at the
electroweak (or higher) scale $\mu_{\rm high}$. Next, one applies the
Renormalization Group Equations (RGE)  
\beq \label{eq:RGE}
\mu \f{d}{d\mu} C_i(\mu) = \gamma_{ji}(\mu) C_j(\mu) \, , 
\eeq
to evolve $C_i(\mu)$ to the relevant low-energy scale $\mu_{\rm
low}$. In this way, large logarithms $\ln(\mu_{\rm high}^2/\mu_{\rm
low}^2)$ are resummed from all orders of the perturbation series.   

Neglecting QED effects, the Anomalous Dimension Matrix (ADM) $\hat
\gamma(\mu)$ has the following perturbative expansion 
\beq \label{eq:ADMexp}
\hat \gamma (\mu) = \sum_{k = 0} \left(\f{\as(\mu)}{4\pi}\right)^{k +
1} \hat \gamma^{(k)} \, ,   
\eeq
where $\as = \gs^2/(4\pi)$ is the strong coupling constant. 

The purpose of our present work is to evaluate the entries
of $\hat \gamma^{(2)}$ that correspond to the $\ord (\as^3)$ strong 
mixing of the dipole operators of \Eq{eq:dipoles} containing quark
fields. These entries can be extracted from the three-loop QCD
renormalization constants in the effective theory. Lower-order entries
are already known from previous calculations \cite{SVZ,Cella:px,
Misiak:1994zw, Broadhurst:1994se, Gambino:2003zm}. The three-loop QCD
self-mixing of the electromagnetic dipole operator coincides with the
anomalous dimension of the rank-two antisymmetric tensor current that
has been calculated in Ref.~\cite{Gracey:2000am}. We confirm all these
findings. Our remaining three-loop results are entirely new.    

The main phenomenological motivation for our work is to provide a new
contribution to the calculation of the $\BtoXsgamma$ branching ratio
at the Next-to-Next-to-Leading Order (NNLO) in QCD. Including these
$\ord (\as^2)$ corrections is necessary to reduce the theoretical
uncertainty of the SM calculation \cite{bsgSM} below the current
experimental one \cite{hfag}. Several steps in this direction have
already been made \cite{Bieri:2003ue, Misiak:2004ew,
Gorbahn:2004my}. Our findings can also be relevant for the CP-odd
electric dipole moment of the neutron, provided the new physics
matching scale is sufficiently high.   

Our calculation completes the three-loop QCD ADM for the whole
dimension-five part of the effective Lagrangian, because the dipole
operators in \Eq{eq:dipoles} are the only EOM-non-vanishing operators
in this sector. Simultaneously, our results establish the whole
three-loop QCD ADM for the $| \Delta F | = 1$ operators of dimensions
five and six that arise in the SM case --- all other three-loop
ADM entries for such operators are known from previous publications 
\cite{Gambino:2003zm, Gorbahn:2004my, Chetyrkin:1996vx}. The only ADM 
entries that remain to be calculated for $\BtoXsgamma$ at the NNLO in
QCD correspond to the four-loop mixing of certain four-quark operators
into the dipole operators \cite{MCz}.   

For definiteness in the further discussion, we shall choose the
flavors, chiralities, normalization and names of the dipole operators
as it is usually done in the phenomenological analyses of
$\BtoXsgamma$, namely    
\beq \label{eq:magneticoperators}
\begin{split}
Q_7 &= \f{e}{16\pi^2}\mb\left(\sbL\sigma^{\mu\nu}\bR\right)F_{\mu\nu} 
\, , \\[2mm]
Q_8 &=
\f{\gs}{16\pi^2}\mb\left(\sbL\sigma^{\mu\nu}T^a\bR\right)G^a_{\mu\nu} 
\, .
\end{split}
\eeq
However, we stress that the $2 \times 2$ ADM which we calculate is the
same for any pair of such quark dipole operators, including the
flavor-conserving ones. There is no mixing between dipole operators 
of different flavor content, even in the flavor-conserving sector
\cite{Misiak:1994zw, Gracey:2000am}.  

In order to remove the divergences of all possible off-shell
one-particle-irreducible (1PI) Green's functions with single
insertions of $Q_7$ and $Q_8$, we have to introduce the following
EOM-vanishing counterterms \cite{Misiak:1994zw, Gambino:2003zm}   
\beq \label{eq:eomvanishingoperators}
\begin{split}
Q_{\smallDsl \hspace{0mm} \smallDsl} &= \f{1}{16\pi^2} \mb\sbL \Dsl
\Dsl \bR \, , \\[2mm]  
Q_{\smallDsl \hspace{0mm} \smallGsl} &= \f{i\gs}{16\pi^2}
\mb \sbL \Big ( \hspace{-0.5mm} \stackrel{\leftarrow}{\Dsl}
\hspace{-0.5mm} \Gsl - \Gsl \Dsl \Big ) \bR \, , 
\end{split}
\eeq
where $D_\mu = \partial_\mu + i \gs G_\mu + i e Q_d A_\mu$ and
$\stackrel{\! \leftarrow}{D_\mu} \hspace{1.5mm} = \hspace{1.5mm}
\stackrel{\! \leftarrow}{\partial_\mu} - i \gs G_\mu \\ - i e Q_d
A_\mu$ denote the covariant derivatives of the gauge group $\SUC
\times U(1)_{\rm em}$ acting on the fields to the right and left, 
respectively, and we have used the definition $G_\mu = G^a_\mu T^a$ 
for the matrix-valued gluon field. $A_\mu$ is the photon field, and
the color generators are normalized so that $\mbox{Tr}(T^a T^b) = 
\delta^{ab}/2$.    

Notice that the operator $Q_{\smallDsl \hspace{0mm} \smallDsl}$ is
gauge-invariant, while $Q_{\smallDsl \hspace{0mm} \smallGsl}$ is 
not. The appearance of such operators is expected on general grounds
\cite{collins,Simma:1993ky}. In principle, one could also encounter
nonphysical counterterms that can be written as
Becchi-Rouet-Stora-Tyutin (BRST) variations of some other operators, 
so-called BRST-exact operators. However, they turn out to be
unnecessary in the case of the dipole operator mixing. This issue is
discussed in more detail in Refs.~\cite{Misiak:1994zw, Gambino:2003zm,
Gorbahn:2004my}.  

{%
\begin{figure}[!t]
\hspace{-4mm}
\includegraphics[width=0.5\textwidth]{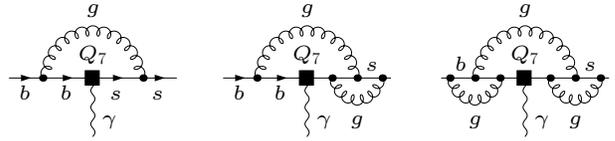}
\caption{Typical examples of 1PI diagrams describing the self-mixing
of $Q_7$ at the one-, two- and three-loop level.}  
\label{fig:mixing}
\end{figure}
}%

We perform the calculation using dimensional regularization and the
$\MSbar$ scheme. As far as the matrix $\gamma_5$ is concerned, its
only relevant property in our case is
$\left[ \gamma_5, \gamma_\mu \gamma_\nu \right ] = 0$,
which, to our knowledge, holds in all the commonly used schemes for the
treatment of $\gamma_5$, including the Naive Dimensional
Regularization (NDR) and t'Hooft-Veltman (HV) schemes. However, the
ADM beyond one loop in the Dimensional REDuction (DRED) scheme is 
different from the one we find here because this scheme does not   
coincide with standard dimensional regularization even in the absence 
of $\gamma_5$. A description of the properties of the NDR, HV and DRED
schemes, as well as a list of relevant original articles can be found
in Ref.~\cite{Buras:1989xd}. 

The necessary three-loop renormalization matrix is found by
calculating the one- and two-loop $\btos$, $\btosgamma$, $\btosgluon$
amputated Green's functions with single insertions of $Q_7$, $Q_8$,
$Q_{\smallDsl \hspace{0mm} \smallDsl}$ and $Q_{\smallDsl \hspace{0mm}
\smallGsl}$, as well as the three-loop $\btosgamma$ and $\btosgluon$
amplitudes with insertions of $Q_7$ and $Q_8$. Sample diagrams are
shown in \Fig{fig:mixing}. The corresponding one-, two- and three-loop
amplitudes are evaluated using the method that has been described in
Refs.~\cite{Misiak:1994zw, Gambino:2003zm, Chetyrkin:1997fm}. We
perform the calculation off shell in an arbitrary $R_\xi$ gauge, which
allows us to explicitly check the gauge-parameter independence of the
mixing among physical operators. To distinguish between infrared and
ultraviolet (UV) divergences, we introduce a common mass $M$ for all
fields, expanding all loop integrals in inverse powers of $M$. This
makes the calculation of the UV divergences possible at three loops,
as $M$ becomes the only relevant internal scale, and three-loop
tadpole integrals with a single nonzero mass are known
\cite{Chetyrkin:1997fm, Broadhurst:1998rz}. On the other hand, this
procedure requires to take into account insertions of the nonphysical
operators $Q_{\smallDsl \hspace{0mm} \smallDsl}$ and $Q_{\smallDsl
\hspace{0mm} \smallGsl}$, as well as of the following counterterm of
dimension three:
\beq \label{eq:gaugevariantoperator}
M^2 \mb \sbL \bR \, .  
\eeq 
A comprehensive discussion of the technical details of the
renormalization of the effective theory and the actual calculation of
the operator mixing is given in Refs.~\cite{Misiak:1994zw,
Gambino:2003zm}. 

Having summarized our method, we now present our results for arbitrary
numbers of down- and up-type quark flavors denoted by $\ND$ and $\NU$,
respectively. The ADM depends on the total number of active quark
flavors $\NF = \NU + \ND$, and their ``total'' electric charge
$\overline{Q} = \NU Q_u + \ND Q_d$. The regularization- and
renormalization-scheme independent matrix $\hat{\gamma}^{(0)}$ is
given by    
\beq \label{eq:gamma0}
\hat{\gamma}^{(0)} = 
\left (
\begin{array}{cc}
{\scriptscriptstyle \f{32}{3}} & 0 \\ {\scriptscriptstyle \f{32}{3}}
{\scriptstyle Q_d} & {\scriptscriptstyle \f{28}{3}}   
\end{array}
\right ) .    
\eeq

While the matrix $\hat{\gamma}^{(0)}$ is renormalization-scheme
independent, $\hat{\gamma}^{(1)}$ and $\hat{\gamma}^{(2)}$ are not. 
In the $\MSbar$ scheme, we obtain 
\beq \label{eq:gamma1}
\hat{\gamma}^{(1)} = 
\left (
\begin{array}{cc}
{\scriptscriptstyle \f{1936}{9} - \f{224}{27} \NF} & 0 \\
{\scriptscriptstyle \left( \f{368}{3} - \f{224}{27} \NF \right)}
{\scriptstyle Q_d} & {\scriptscriptstyle \f{1456}{9} - \f{61}{27} \NF}
\end{array}
\right ) , 
\eeq
and  
\begin{align} \label{eq:gamma2}
\hat{\gamma}^{(2)} & =
\left (
\begin{array}{cc}
{\scriptscriptstyle -\f{1856}{27} -\f{1280}{9}\NF} & 0 \\
{\scriptscriptstyle \f{640}{9}} {\scriptstyle \overline{Q}}
{\scriptscriptstyle - \left( \f{1856}{27} + \f{1280}{9}\NF \right)}
{\scriptstyle Q_d} & {\scriptscriptstyle -\f{28624}{27} -
\f{1312}{9}\NF}   
\end{array}
\right ) \zetathree \non \\[2mm]
& \hspace{-7.25mm}
+\left(\!\!
\begin{array}{cc} 
{\scriptscriptstyle \f{307448}{81} - \f{23776}{81}\NF - \f{352}{81}
\NF^2} & 0 \\ {\scriptscriptstyle -\f{1600}{27}} {\scriptstyle
\overline{Q}} {\scriptscriptstyle + \left( \f{159872}{81} -
\f{17108}{81}\NF - \f{352}{81}\NF^2 \right)} {\scriptstyle Q_d} & 
{\scriptscriptstyle \f{268807}{81} - \f{4343}{27}\NF -
\f{461}{81}\NF^2}  
\end{array}
\!\!\right) . \non \\[1mm]
\end{align}
%
%\begin{widetext}
%\beq \label{eq:gamma2}
%\hat{\gamma}^{(2)} = 
%\left (
%\begin{array}{cc}
%{\scriptscriptstyle \f{307448}{81} - \f{23776}{81}\NF - \f{352}{81}
%\NF^2 - \left ( \f{1856}{27} + \f{1280}{9}\NF \right ) \,
%{\scriptstyle \zetathree}} & 0 \\ 
%{\scriptscriptstyle -\f{1600}{27} {\scriptstyle \overline{Q}} + \left(
%\f{159872}{81} - \f{17108}{81}\NF - \f{352}{81}\NF^2 \right)
%{\scriptstyle Q_d} + \left ( \f{640}{9} {\scriptstyle \overline{Q}} -
%\left( \f{1856}{27} + \f{1280}{9}\NF \right) {\scriptstyle Q_d} \right
%) \, {\scriptstyle \zetathree}} & {\scriptscriptstyle \f{268807}{81} -
%\f{4343}{27}\NF - \f{461}{81}\NF^2 - \left ( \f{28624}{27} +
%\f{1312}{9}\NF \right ) \, {\scriptstyle \zetathree}}    
%\end{array}
%\right ) \, . 
%\eeq
%\end{widetext}
%
We remark that the explicit electric charge $Q_d$ originates solely
from the quarks in the operators, and thus has to be replaced by $Q_u$
for operators containing up-quark fields. As it is characteristic for
three-loop anomalous dimensions, the entries of $\hat{\gamma}^{(2)}$
involve terms proportional to the Riemann zeta function $\zetathree
\approx 1.20206$.   

Of course, the presence of the bottom quark mass in the normalization
of the dipole operators of \Eq{eq:magneticoperators} affects the
values of $\hat{\gamma}^{(k)}$. Had we decided to define the operators
without quark mass in their normalization, the results in
Eqs.~(\ref{eq:gamma0}) to (\ref{eq:gamma2}) would need to be replaced
by $\gamma_{ij}^{(k)} - \gamma_m^{(k)} \delta_{ij}$, where  
\beq \label{eq:gammams}
\begin{split}
\gamma_m^{(0)} & = 8 \, , \\[2mm]
\gamma_m^{(1)} & = \f{404}{3} - \f{40}{9} \NF \, , \\
\gamma_m^{(2)} & = 2498 - \left(\f{4432}{27} + \f{320}{3} \zetathree 
\right) \NF - \f{280}{81} \NF^2 \, ,  
\end{split}
\eeq
are the expansion coefficients of the quark mass anomalous
dimension. In particular, verifying that our result for
$\gamma_{77}^{(2)}$ is in agreement with Eq.~(8) of
Ref.~\cite{Gracey:2000am} requires to perform such a replacement
because no quark mass was present in the normalization of the tensor 
current considered there. We note that there is a misprint in the last
line of Eq.~(7) of the latter paper. Obviously, the factor $C_F^2$
should read $N_f^2$.

The RGE for the Wilson coefficients given in \Eq{eq:RGE}, has the
following general solution 
\beq \label{eq:evol}
C_i(\mub) = U_{ij}(\mub,\muw) C_j(\muw) \, ,
\eeq
where the matching and the low-energy scales have been denoted by
$\muw$ and $\mub$, respectively. In the $\BtoXsgamma$ case, one has
$\muw = \ord (\MW)$ and $\mub = \ord (\mb)$. The evolution matrix 
$U_{ij}(\mub,\muw)$ depends on the strong gauge coupling ratio $\eta =
\as(\muw)/\as(\mub)$. The ADM that we have calculated allows us to
find the complete $\ord(\as^2)$ contributions to  
\beq \label{eq:deltac7}
\Delta C_7(\mub) = \sum_{j = 7, 8} U_{7j} (\mub,\muw) C_j(\muw) \, , 
\eeq
and
\beq \label{eq:deltac8}
\Delta C_8(\mub) = U_{88} (\mub,\muw) C_8(\muw) \, .
\eeq
Denoting these $\ord(\as^2)$ contributions by $\Delta^{(2)}
C_{7}(\mub)$ and $\Delta^{(2)} C_{8}(\mub)$, respectively, and using
the general NNLO formalism presented in Refs.~\cite{Gorbahn:2004my,
NNLOmatrixkernel}, we obtain for $\NF = 5$, $Q_d = -1/3$ and
$\overline{Q} = 1/3$:    
\begin{align} \label{eq:nnlophotonic}
& \Delta^{(2)} C_7(\mub) = \left ( \f{\as(\mub)}{4 \pi} \right
)^2 \Bigg [ \eta^{\f{62}{23}} C_7^{ (2)} (\muw) \non \\ 
& + \text{\scalebox{1.0}{$\f{8}{3}$}} \left ( \eta^{\f{60}{23}} -
\eta^{\f{62}{23}} \right ) C_8^{ (2)} (\muw) \non \\[2mm]        
& - \text{\scalebox{1.0}{$\f{37208}{4761}$}} \left (
\eta^{\f{39}{23}} - \eta^{\f{62}{23}} \right ) C_7^{ (1)}
(\muw) \non \\[2mm]
& - \left ( \text{\scalebox{1.0}{$\f{7164416}{357075}$}} 
\eta^{\f{37}{23}} - \text{\scalebox{1.0}{$\f{297664}{14283}$}}
\eta^{\f{39}{23}} \right. \non \\[-4mm]
\\
& \left. - \text{\scalebox{1.0}{$\f{256868}{14283}$}} 
\eta^{\f{60}{23}} + \text{\scalebox{1.0}{$\f{6698884}{357075}$}}
\eta^{\f{62}{23}} \right ) C_8^{ (1)} (\muw) \non \\[1mm] 
& + \left ( 16.6516 \, \eta^{\f{16}{23}} - 61.0768 \,
\eta^{\f{39}{23}} + 44.4252 \, \eta^{\f{62}{23}} \right ) C_7^{(0)}
(\muw) \non \\   
& + \left ( 36.4636 \, \eta^{\f{14}{23}} - 44.4043 \,
\eta^{\f{16}{23}} - 135.3141 \, \eta^{\f{37}{23}} \right. \non
\\[-1mm]   
& \left. + 146.6159 \, \eta^{\f{39}{23}} + 15.4051 \,
\eta^{\f{60}{23}} - 18.7662 \, \eta^{\f{62}{23}} \right ) C_8^{ (0)} 
(\muw) \Bigg ] \, , \non \\
& \text{and} \non \\
\label{eq:nnlogluonic}
& \Delta^{(2)} C_8^{} (\mub) = \left ( \f{\as(\mub)}{4 \pi} \right )^2 
\Bigg [ \eta^{\f{60}{23}} C_8^{ (2)} (\muw) \non \\
& - \text{\scalebox{1.0}{$\f{64217}{9522}$}} \left ( \eta^{\f{37}{23}}
- \eta^{\f{60}{23}} \right ) C_8^{ (1)} (\muw) \\  
& + \left ( 39.7055 \, \eta^{\f{14}{23}} - 45.4824 \,
\eta^{\f{37}{23}} + 5.7769 \, \eta^{\f{60}{23}} \right ) C_8^{ (0)} 
(\muw) \Bigg ] \, . \non 
\end{align}
Explicit expressions for the relevant $b\to s \gamma$ and $b\to s g$
matching conditions
\beq
C_i(\muw) = \sum_{k = 0} \left(\f{\as(\muw)}{4\pi}\right)^k
C_i^{(k)}(\muw) \, , 
\eeq
can be found in Ref.~\cite{Misiak:2004ew}.

Setting $\as (\MZ) = 0.118$, $\mt = \mt (\mt) = 168.5 \GeV$, $\muw =
\MW = 80.425 \GeV$ and $\mub = 4.8 \GeV$, one obtains
\beq \label{eq:dc7}
\Delta^{(2)} C_7(\mub) \approx \left(\f{\as(\mub)}{4\pi}\right )^2
17.2 \, \approx \, 0.0051 \, ,
\eeq
and
\beq \label{eq:dc8}
\Delta^{(2)} C_8(\mub) \approx \left(\f{\as(\mub)}{4\pi}\right )^2
6.4 \, \approx \, 0.0019 \, .   
\eeq
The correction in \Eq{eq:dc7} causes a suppression of the
$\BtoXsgamma$ branching ratio by around $3 \%$. However, one should
bear in mind that the size of the correction depends strongly on the
exact value chosen for $\muw$. Only the {\em total} values of
$C_i(\mub)$ are guaranteed to become less $\muw$-dependent once more
orders in their perturbative expansion are being included. At the
moment, no complete $\ord(\as^2)$ expression for $C_7(\mub)$ is
available because one and only one element on the right hand side of
\Eq{eq:evol} remains unknown at this order, namely the element $U_{72}
(\mub,\muw)$ of the evolution matrix that corresponds to the
current-current operator $Q_2 = (\sbL \gamma_\mu \cL)(\cbL \gamma^\mu
\bL)$. It is going to become available once the calculation of the
four-loop mixing of the relevant four-quark into the dipole operators
is completed \cite{MCz}.\\     

To conclude: We have evaluated the complete three-loop $\ord (\as^3)$
mixing among the dipole operators, that is, in the whole
dimension-five sector of the effective theory that describes processes
occurring much below the electroweak scale. Our results are
particularly relevant for the NNLO analysis of radiative $B$ decays in
the SM and many of its extensions.\\ 

\noindent {\bf Acknowledgments:}
We would like to thank Matthias Steinhauser for providing us with an updated
version of {\tt MATAD} \cite{Steinhauser:2000ry}. Our calculations made
extensive use of the Fermilab General-Purpose Computing Farms
\cite{Albert:2003vv}. M.~G.\ and U.~H. \ appreciate the warm hospitality of
the Institute T31 for Theoretical Elementary Particle Physics at the Technical
University of Munich where part of this work has been performed. The work of
U.~H.\ is supported by the U.S. Department of Energy under contract
No.~DE-AC02-76CH03000. M.~M.\ was supported in part by the Polish Committee
for Scientific Research under the grant 2~P03B~078~26, and from the European
Community's Human Potential Programme under the contract HPRN-CT-2002-00311,
EURIDICE.

\end{document}